\documentclass[12pt,a4paper]{article}
\usepackage{amsmath}
\usepackage{graphicx,amsmath,amssymb}

\usepackage{cite}
\usepackage{graphicx,amsmath,amssymb}

\DeclareMathAlphabet{\mathpzc}{OT1}{pzc}{m}{it}

\title{Double-logarithms in Einstein-Hilbert gravity and supergravity} 
\author{Jochen Bartels$^{1}$, Lev N. Lipatov$^{1,2}$, Agust{\' \i}n Sabio Vera$^{3,4}$\\ 
\\
$^1$ II. Institute of Theoretical Physics, Hamburg U., Germany\\
$^2$ St. Petersburg Nuclear Physics Institute, Russia\\
$^3$ Instituto de F{\' \i}sica Te{\' o}rica UAM/CSIC, Nicol{\'a}s Cabrera 15\\ 
\& U. Aut{\' o}noma de Madrid, E-28049 Madrid, Spain\\
$^4$ CERN,  Geneva, Switzerland} 

\begin{document} 

\maketitle 

We study the interplay between graviton reggeization and double-logarithmic in energy contributions to  
four-graviton scattering in theories with and without supersymmetry. Predictions to all orders 
in the gravitational coupling are given for these double-logarithms. As the number of supersymmetries grows 
these terms generate a convergent behaviour for the amplitudes at very high energies. At two-loop level, we find agreement with previous 
exact results for ${\cal N}=8$ supergravity and with those of Boucher-Veronneau and Dixon, who studied 
the ${\cal N}=4,5,6$ supergravities using a conjectured double-copy structure of gravity.

\section{Introduction}

Collider phenomenology and the anti de Sitter / conformal field theory (AdS/CFT) correspondence~\cite{Maldacena:1997re,Gubser:1998bc,Witten:1998qj} have motivated numerous investigations of scattering amplitudes in 
gauge and gravitational theories in recent years. In this context, the study of the ${\cal N} = 4$ supersymmetric Yang-Mills theory has been particularly successful since it is possible to investigate loop corrections by only evaluating a reduced set of master topologies~\cite{Bern:2010tq}. These results for ${\cal N}=4$ SUSY can then be used to also obtain amplitudes in 
${\cal N} = 8$ supergravity, where the set of minimal topologies is still valid, and to address the question of the renormalizability of the theory at higher orders~\cite{Bern:1998ug,Bern:2009kd}. This procedure is based on a 
conjectured double-copy structure of gravity~\cite{Bern:2002kj} which has recently been applied to ${\cal N} = 4,5,6$ supergravities at 
two-loops~\cite{BoucherVeronneau:2011qv}. In Einstein-Hilbert gravity progress is slower~\cite{Donoghue:1994dn,BjerrumBohr:2002kt,Donoghue:1999qh} since supersymmetry or string theory based techniques~\cite{Dunbar:1994bn,Bern:1993wt,Bern:1991an} cannot help.

It is also possible to get important information 
of the all-orders structure of scattering amplitudes when they are considered in certain kinematical regions. 
An interesting example is the study of graviton scattering in multi-Regge kinematics (MRK). In this case the amplitudes present a factorized form which can be interpreted in terms of the exchange of reggeized gravitons~\cite{Grisaru:1975tb,Grisaru:1981ra}  
 together with eikonal and double-logarithmic terms~\cite{Lipatov:1982vv,Lipatov:1982it,Lipatov:1991nf}. 
 These contributions, together with new interaction vertices, can be described by means of a high energy effective action~\cite{Lipatov:2011ab}. It is noteworthy that the graviton emission vertex in MRK can be written as a double copy of the corresponding~\cite{BFKL1,BFKL2,BFKL3} QCD gluon emission vertex~\cite{Lipatov:1982vv,Lipatov:1982it,Lipatov:1991nf,SabioVera:2011wy}.

In the present work double-logarithmic in energy contributions to four-graviton scattering to all orders 
in the gravitational coupling will be evaluated. This will be done for arbitrary supergravities as well as for Einstein-Hilbert gravity. We will improve previous results based on the resummation of ladder-like diagrams by considering the full set of contributing topologies. The truncation of our results to two loops is in agreement with recent calculations in the literature for 
${\cal N} = 4, 5, 6, 8$ supergravities. The all-orders resummation of these contributions generates amplitudes which grow with energy when ${\cal N}<4$ and asymptotically tend to zero 
when ${\cal N}>4$. ${\cal N}=4$ supergravity corresponds to a critical theory where the leading 
double-logarithmic contributions cancel.

\section{Double-logarithmic approximation}

For our analysis it is convenient to define the following normalization for the four-point amplitudes
\begin{eqnarray}
{\cal A}_{4,(N)} &=& {\cal A}_4^{\rm Born} {\cal M}_{4,(N)},\\
{\cal A}_4^{\rm Born} &=& \kappa^2 \frac{s^3}{t u},\\
{\cal M}_{4,(N)} &=& 1 + \sum_{L=1}^\infty {\cal M}_{4,(N)}^{(L)},
\label{eqnnotation}
\end{eqnarray}
where $L$ corresponds to the loop order and $N$ labels the number of gravitinos in the theory. The Mandelstam invariants are 
$s=(p_1+p_2)^2$, $t=(p_1-p_3)^2$ and $u=(p_1-p_4)^2$. $\kappa^2 = 8 \pi G$, with $G$  being the Newton's constant. 

\subsection{One-loop amplitudes}

When the one loop amplitude ${\cal M}_{4,(N)}^{(1)}$ is calculated in the Regge limit (with $s \gg -t=|q|^2$) the graviton Regge trajectory~\cite{Lipatov:1982vv,Lipatov:1982it,Lipatov:1991nf} 
\begin{eqnarray}
\omega (q) &=&\nonumber\\
&&\hspace{-1.3cm}\frac{\alpha |q|^2}{\pi}\int \frac{d^2k }{|k|^2|q-k|^2}
\left(\frac{(\vec{k},\vec{q}-\vec{k})^2}{|k|^2}+
\frac{(\vec{k},\vec{q}-\vec{k})^2}{|q-k|^2}-|q|^2+\frac{N}{2}\,(\vec{k},\vec{q}-\vec{k})\right)
\end{eqnarray}
appears multiplied by  $\ln{(s/|q|^2)}$. We have used the notation $\alpha = \kappa^2 / (8 \pi^2)$. This expression contains 
both infrared and ultraviolet divergencies which can be regulated by, respectively, the cut-offs $\lambda$ and  
$\Lambda$, to obtain
\begin{equation}
\omega (q) =-\alpha \,|q|^2
\,\left(\ln \frac{ |q|^2}{\lambda ^2}+\frac{N-4}{2}\,\ln \frac{\Lambda^2}{|q| ^2}\right)\,.
\label{trajgrav}
\end{equation}
The ultraviolet divergence at $\Lambda \to \infty$ is not fundamental because gravity
is renormalizable at one loop.  It has a kinematical origin which means that the parameter
$\Lambda$ should be substituted by $\sqrt{s}$, leading to the
appearance of the double-logarithmic term $\sim \alpha \ln ^2s$ in the elastic scattering
amplitude. 

To understand this point in more detail, let us recall that the one loop contribution at high energies can be obtained making use of the Sudakov parametrization for the virtual particle momentum,
\begin{equation}
k=\beta \,p_1+\alpha \,p_2+k_\perp \,,\,\,d^4k=\frac{|s|}{2}\,d\alpha \,d\beta \,d^2k_\perp \,,
\end{equation}
where $p_1$ and $p_2$ are the momenta of the colliding particles.
Calculating the Feynman integral over $\alpha$ by residues we obtain with leading logarithmic
accuracy the following expression
 \begin{equation}
{\cal A}_4^{(1)} (s,t)\sim \frac{1}{\pi }\int \frac{d^2k_\perp}{|k_\perp|^2}\,
\int ^s_{|q|^2} \frac{d (\beta s)}{\beta s -|k_\perp|^2}\,.
\label{brems}
\end{equation}
It is then clear that the ultraviolet divergence at $|k_\perp|\rightarrow \infty$ is absent. In the infrared region
of integration $|k^2_\perp|\ll |q|^2$ the above expression factorizes in the form
${\cal A}_4^{(1)} \sim \ln (|q|^2/\lambda ^2) \, \ln (s/|q|^2)$.
This Regge factor containing the infrared divergence can be extracted to all
orders in perturbation theory and the amplitude with double-logarithmic accuracy can be  conveniently presented using the following Mellin transform:
\begin{equation}
{\cal A}_{4,(N)} (s,t) = {\cal A}_4^{\rm Born}\,\left(\frac{s}{|q|^2}\right)^{-\alpha |q|^2\,\ln \frac{ |q|^2}{\lambda ^2}}
\int _{\delta-i\infty}^{\delta +i\infty}\frac{d\,\omega }{2\pi i}\,
\left(\frac{s}{|q|^2}\right)^\omega \frac{ f ^{(N)}_\omega}{\omega} \,,\,\,\delta>0\, ,
\label{factor}
\end{equation}
where the tree level amplitude is ${\cal A}_4^{\rm Born} \simeq \kappa^2 s^2 / |q|^2$ and
the integral over $\omega$ contains the contribution of the virtual gluons and gluinos only with $|k_\perp|^2>|q|^2$. The $t$-channel partial wave $f ^{(N)}_\omega$ in
the double-logarithmic approximation can be  expanded order by order in perturbation theory,
\begin{equation}
f^{(N)}_\omega =\sum _{n=0}^\infty c_n^{(N)}\,\left(\frac{b}{\omega^2}\right)^n\,,
\end{equation}
where $b$ is the dimensionless parameter
\begin{equation}
b=\alpha |q|^2\, .
\end{equation}

In Ref.~\cite{Lipatov:1982vv} one of the authors of this work (L.N.L.) made the assumption 
that double-logarithmic
contributions in gravity appear only from ladder diagrams, which allowed him to obtain
a closed expression for the scattering amplitude in terms of a Bessel function. However, 
even in simpler field theories like QED and QCD there is another source of 
double-logarithmic terms. For example, in $e^+e^-$ backward scattering the diagrams with
virtual soft photons emitted and absorbed by the external fermions are essential. These contributions contain, apart from the integration region $|k_\perp|^2>|q|^2$, the universal infrared divergencies from the region $|k_\perp|^2\ll |q|^2$ in the form of Eq.(\ref{brems}). It is then natural to realize that similar non-ladder contributions also play a role in gravity. At one loop they would lead to the following correction to the elastic
amplitude
\begin{eqnarray}
\frac{{\cal A}^{(1)}_{4,{\rm soft}}}{{\cal A}_4^{\rm Born}} &=&
-b\int _{\lambda ^2}^{s}\frac{d|k_\perp|^2}{|k_\perp|^2}\,
\int ^s_{|q|^2} \frac{d (\beta s)}{\beta s -|k_\perp|^2}\nonumber\\
&=&-b\,\ln \frac{s}{|q|^2}
\left(\ln \frac{|q|^2}{\lambda ^2}+\frac{1}{2}\,\ln \frac{s}{|q|^2}\right),
\end{eqnarray}
which, together with the corresponding ladder correction (including gravitino contributions), 
\begin{equation}
\frac{{\cal A}^{(1)}_{4,{\rm ladder}}}{{\cal A}_4^{\rm Born}}=-b \left(\frac{N-6}{4}\right)\ln^2 \frac{s}{|q|^2}\,
\label{ladd}
\end{equation}
generate the complete one loop correction (cf. Eq.~(\ref{trajgrav})):
\begin{equation}
\frac{{\cal A}^{(1)}_{4,(N)}}{{\cal A}_4^{\rm Born}}=-b\,\ln \frac{s}{|q|^2}\,
\left(\ln \frac{|q|^2}{\lambda ^2}+\left(\frac{N-4}{4}\right)\ln \frac{s}{|q|^2}\right)\,.
\label{resultoneloop}
\end{equation}

To better understand this result let us now compare it to the exact one-loop amplitudes 
available in the literature. In Ref.~\cite{BoucherVeronneau:2011qv} Boucher-Veronneau 
and Dixon used a conjectured double-copy structure of gravity to evaluate 
four-point scattering amplitudes at two loops in ${\cal N}=4,5,6$ supergravities. They also 
calculated again the well-known ${\cal N}=8$ case. 

For ${\cal N}=8$ it is possible to write the exact one-loop amplitude (using $s+t+u=0$) in the form
\begin{eqnarray}
{\cal M}^{(1)}_{4, (N=8)}  &=& \underbrace{\alpha \, t \ln{\left(\frac{-s}{-t}\right)}\ln{\left(\frac{-u}{-t}\right)}}_{\rm Double~Logs} \nonumber\\
&+& \underbrace{\alpha \, \frac{t}{2} \ln{\left(\frac{-t}{\lambda^2}\right)}
\left(\ln{\left(\frac{-s}{-t}\right)}+\ln{\left(\frac{-u}{-t}\right)}\right)}_{\rm Trajectory}  \nonumber\\
&-& \underbrace{\alpha \frac{(s-u)}{2} \ln{\left(\frac{-t}{\lambda^2}\right)}\ln{\left(\frac{-s}{-u}\right)}}_{\rm Eikonal}.
\end{eqnarray}
We have indicated the terms which will generate, in the Regge limit ($s \simeq -u$), the double-logarithms, 
the graviton trajectory and the eikonal pieces.

Following Ref.~\cite{BoucherVeronneau:2011qv} it is possible to relate the previous expression to the  ${\cal N}=4,5,6$ supergravity amplitudes, {\it i.e.}
\begin{eqnarray}
{\cal M}^{(1)}_{4, (N=4)} &=& {\cal M}^{(1)}_{4, (N=8)} + \alpha \, t  \frac{1}{2}\frac{u}{s}
\Bigg\{\left(2-\frac{u\, t}{s^2} \right)\left(\ln^2{\left(\frac{-u}{-t}\right)}+\pi^2\right) \nonumber\\
&+& 1 + \left(\frac{s-u}{s}\right) \ln{\left(\frac{-t}{s}\right)}
+ \left(\frac{u-t}{s}\right) \ln{\left(\frac{-u}{s}\right)}\Bigg\}, \\
{\cal M}^{(1)}_{4, (N=5)} &=& {\cal M}^{(1)}_{4, (N=8)}+ \alpha \, t \frac{3}{4}  \frac{u}{s}
\left(\ln^2{\left(\frac{-u}{-t}\right)}+\pi^2\right), \\
{\cal M}^{(1)}_{4, (N=6)}  &=& {\cal M}^{(1)}_{4, (N=8)} + \alpha \, t \frac{1}{2} \frac{u}{s} 
\left(\ln^2{\left(\frac{-u}{-t}\right)}+\pi^2\right).  
\end{eqnarray}
From the work of Dunbar and Norridge in Ref.~\cite{Dunbar:1994bn} 
we also know the one-loop amplitude in Einstein-Hilbert gravity ($N=0$): 
\begin{eqnarray}
{\cal M}^{(1)}_{4, (N=0)} &=& 
{\cal M}^{(1)}_{4, (N=8)} + \alpha \, t \frac{1}{2}  \frac{u}{s}
{\cal G} \left(\frac{-t}{s}\right),
\end{eqnarray}
with
\begin{eqnarray}
{\cal G}(x) &=& \left(4-10 x+2 x^2 +15 x^3-5 x^4-3 x^5+x^6\right) \nonumber\\
&\times& \Bigg\{\ln^2{(x)}+2 x \ln{(x)}+\pi^2+\sum_{n=2}^\infty 
x^n \left(\frac{2}{n} \ln{(x)} + \sum_{l=1}^{n-1}\frac{1}{l(n-l)}\right)\Bigg\}
\nonumber\\
&+& \left(\frac{341}{30}-\frac{437}{30}x-\frac{47}{2}x^2+\frac{37}{3}x^3
+5 x^4-2 x^5\right) \left(\ln{(x)}+ \sum_{n=1}^\infty \frac{x^n}{n}\right)  \nonumber\\
&+& \frac{961}{90}+\frac{97}{12}x-\frac{85}{12}x^2-2 x^3+x^4.
\end{eqnarray}
Using these expressions we can see that the double-logarithmic contributions to these amplitudes can be written in the form
\begin{eqnarray}
{\cal M}^{(1),{\rm DL}}_{4, (N)}  &=&\left(\frac{N-4}{2}\right) \left(\frac{\alpha \, t}{2}\right) \ln^2{\left(\frac{s}{-t}\right)},
\end{eqnarray}
in agreement with our result in Eq.~(\ref{resultoneloop}).

\subsection{All-loop amplitudes}

Since the sources for the double-logarithmic contributions in gravity are the same as in gauge theories (see Ref.~\cite{Kirschner:1982qf,Kirschner:1982xw,Kirschner:1983di} for 
related calculations in QED and QCD), it is possible to write a similar infrared evolution equation for the partial wave introduced in Eq.~(\ref{factor}), 
namely,
\begin{equation}
f^{(N)}_\omega =1+b\frac{d}{d\,\omega}\,\frac{f^{(N)}_\omega}{\omega}
-b \left(\frac{N-6}{2}\right) \frac{{f^{(N)}_\omega}^2}{\omega ^2},
\label{evoleq}
\end{equation}
which, in pictorial terms, stems from the equation 
\begin{eqnarray}
 \hspace{-1.6cm}\parbox{15mm}{\includegraphics[width=2.2cm,angle=0]{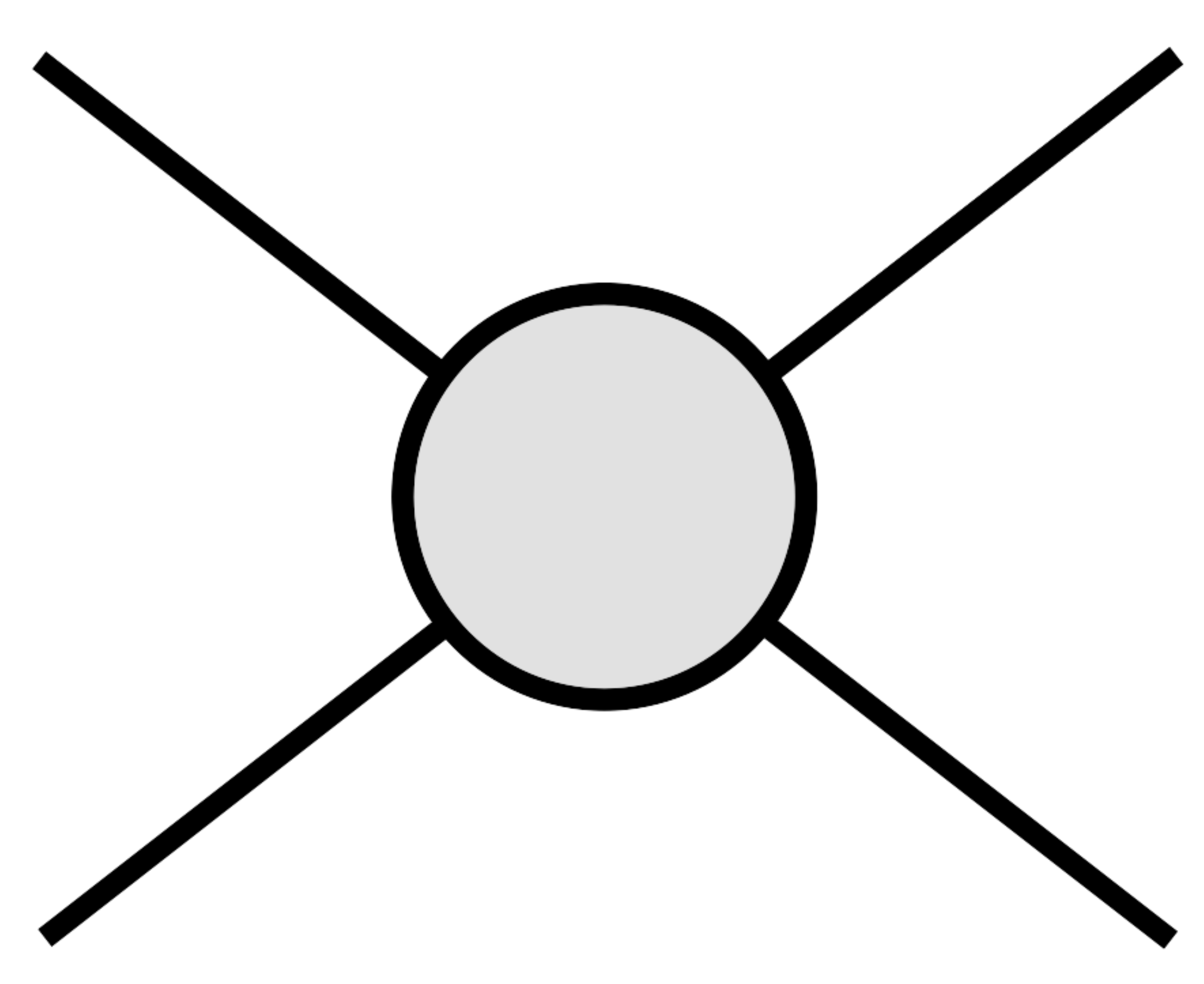}} \hspace{.4cm}
 &=& \hspace{-.3cm} \parbox{15mm}{\includegraphics[width=2.2cm,angle=0]{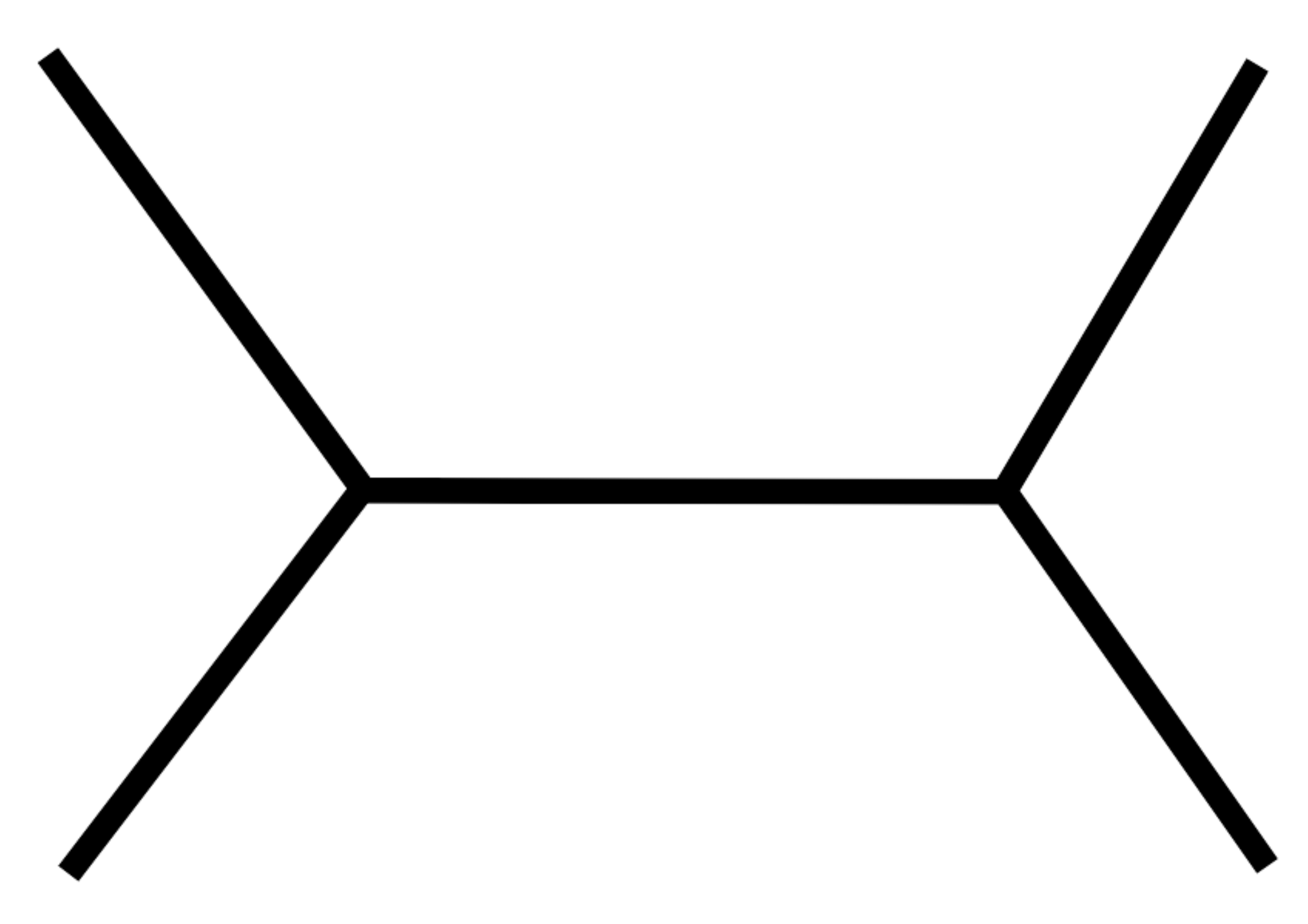}} 
 \hspace{0.6cm}
 + 2 \hspace{.cm} \parbox{15mm}{\includegraphics[width=2.2cm,angle=0]{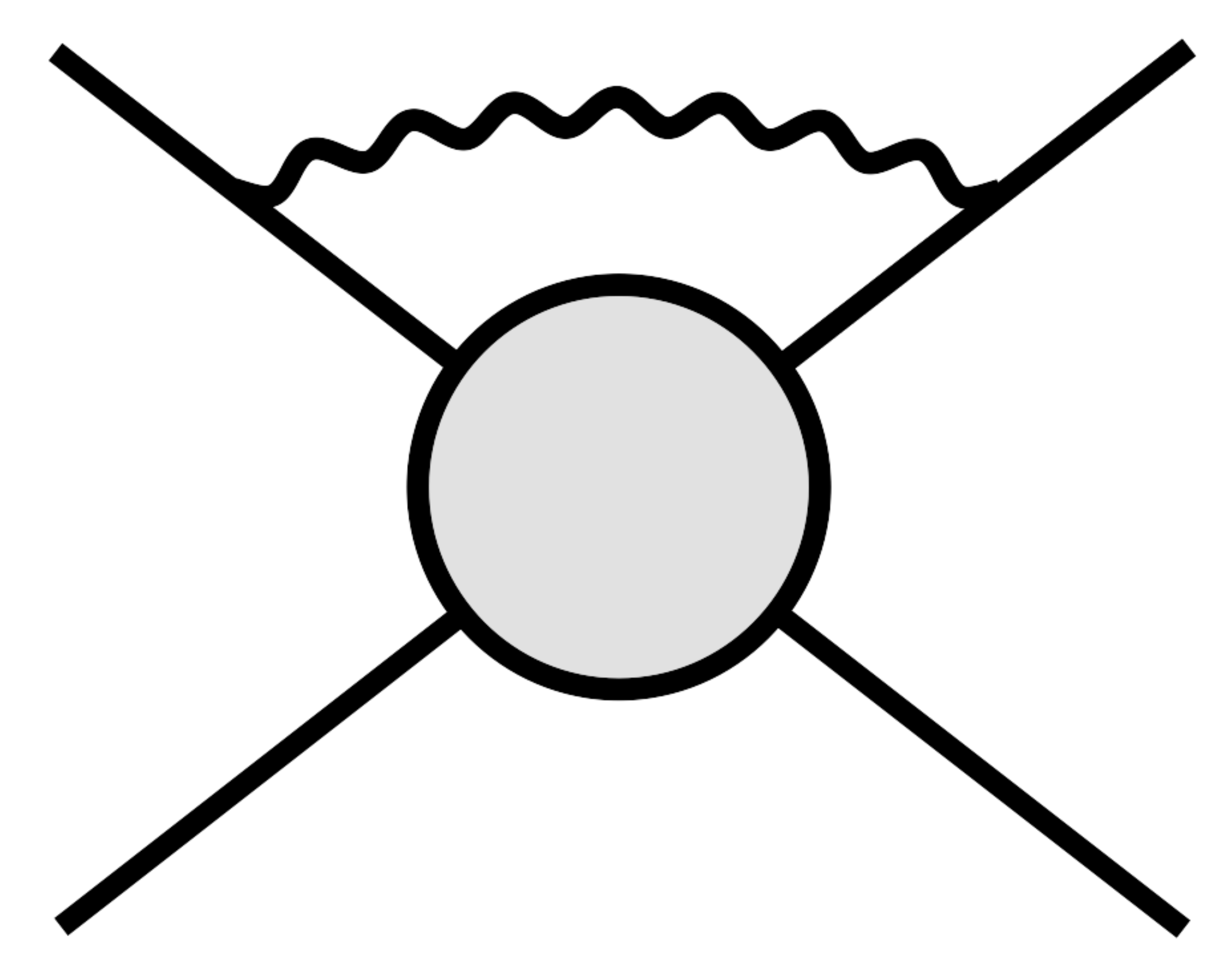}}
 \hspace{0.5cm}
 + 2 \hspace{.cm} \parbox{15mm}{\includegraphics[width=2.2cm,angle=0]{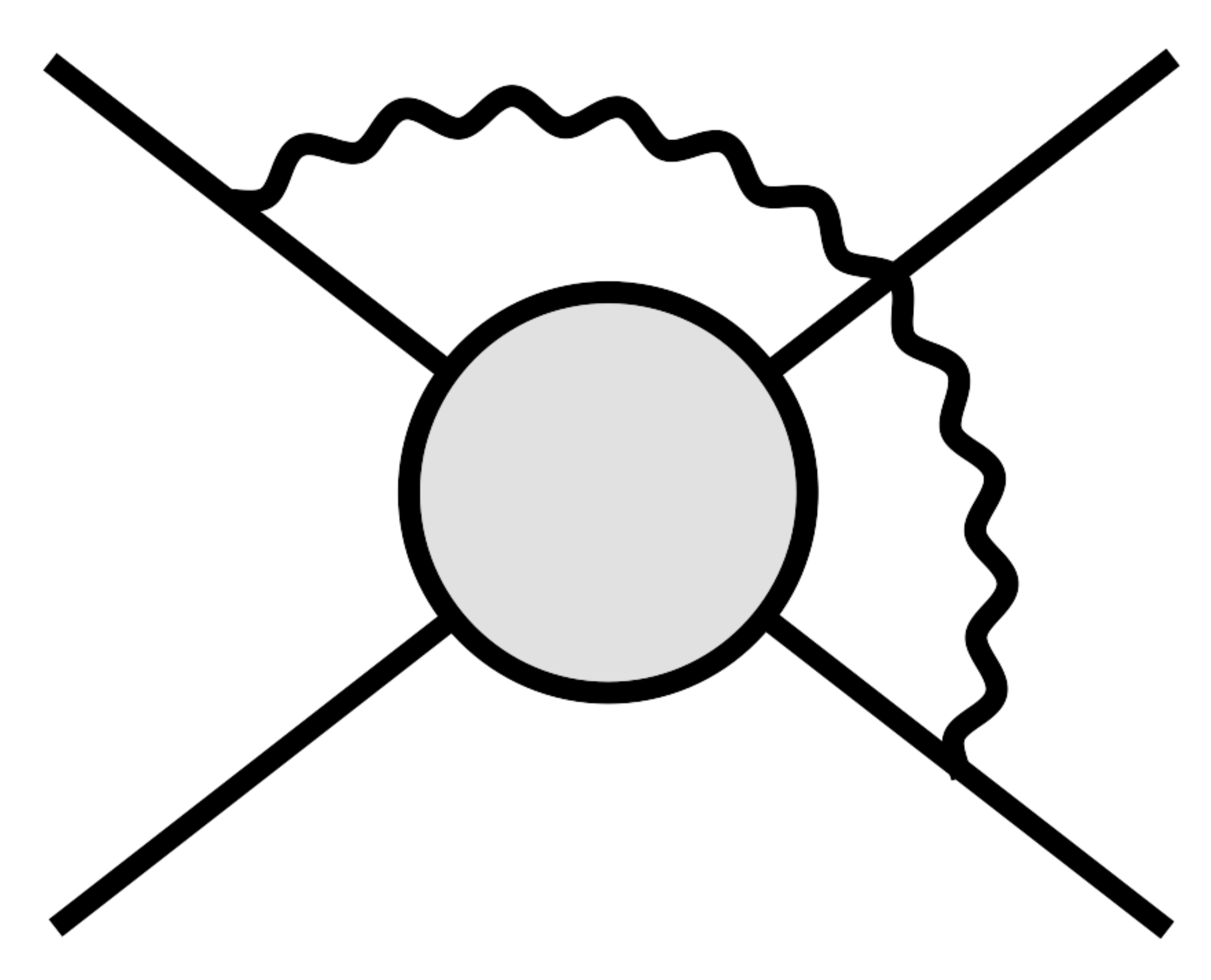}}
 \hspace{0.5cm}
  + \hspace{-.1cm} \parbox{15mm}{\includegraphics[width=2.6cm,angle=0]{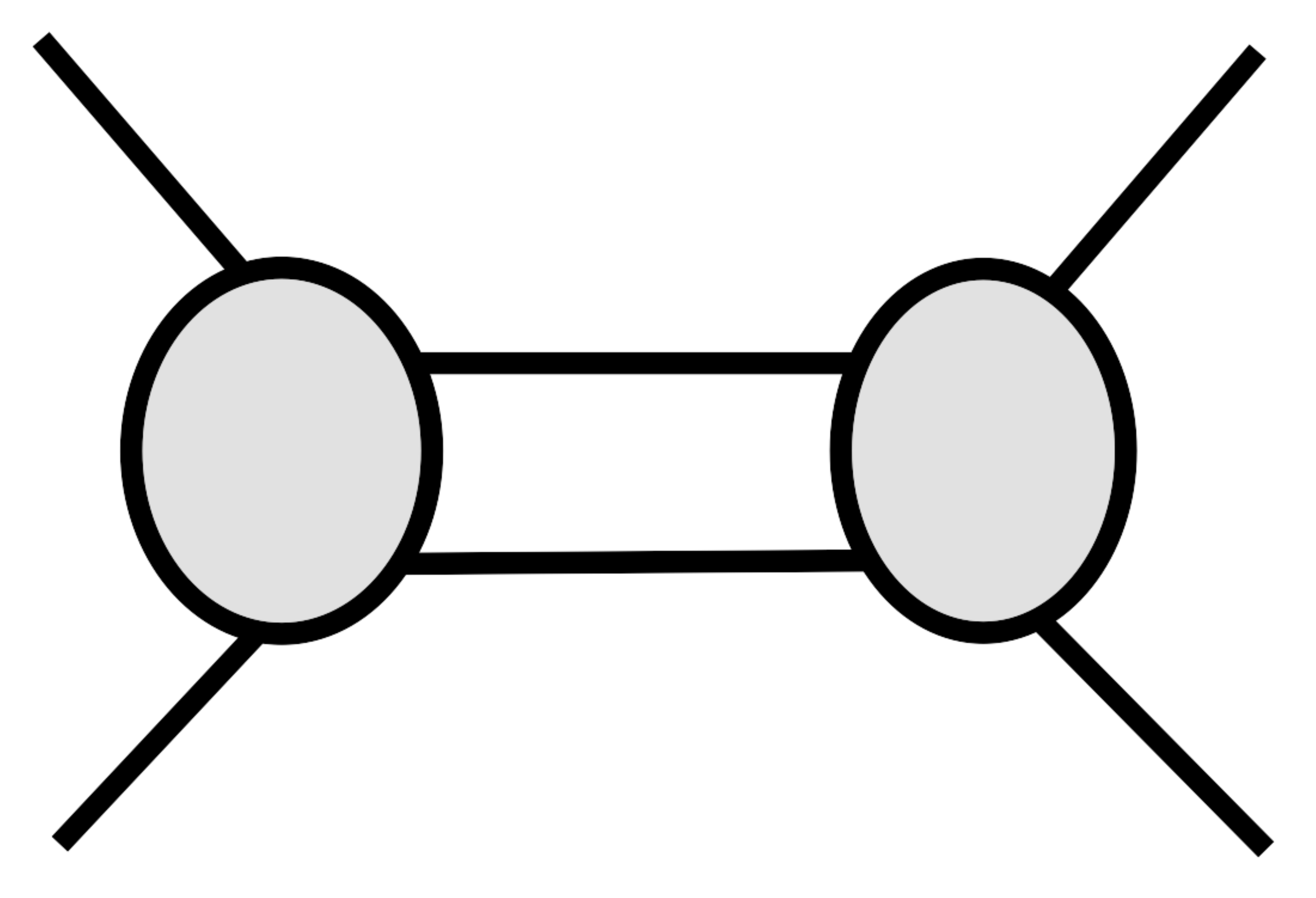}}\hspace{1cm}.\hspace{-.8cm}
\end{eqnarray}
At the right hand side of Eq.~(\ref{evoleq}) the first term proportional to $b$ describes the contribution of the virtual
graviton with the smallest value of $k_\perp$. The second (ladder) term describes the contribution from the pair with the two softest gravitons or gravitinos exchanged in the $t$-channel. Let us indicate that, generally, the emission of the
virtual soft graviton with transverse momentum $k_\perp$ changes the momentum transfer ($q\rightarrow q-k$) for the basic scattering process, and hence modifies the power factors $|q|^{2n}\rightarrow |q-k|^{2n}$ of the corresponding amplitude in each order of perturbation theory. 
Nevertheless, it is correct to neglect the corrections $\sim k_\perp$
because these terms cancel the logarithmic contribution appearing from the integration over  $k_\perp$. 

It is important to indicate that the coefficients of Eq.~(\ref{evoleq}) in front of the two terms proportional to $b$ are chosen in such a way as to reproduce the one loop contribution  calculated in Eq.~(\ref{resultoneloop}). The perturbative solution of the infrared evolution equation~(\ref{evoleq}) has the form
\begin{eqnarray}
f_\omega^{(N)} &=& 1 - \frac{b (N-4)}{2 w^2} + \frac{b^2 (N-4) (N-3)}{2 w^4} \nonumber\\
    &-& \frac{b^3 (N-4) \left(5 N^2- 26 N+36\right)}{8 w^6}\nonumber\\
    &+&\frac{b^4 (N-4)   \left(7 N^3-47 N^2+118 N-108\right)}{8 w^8}\nonumber\\
     &-&\frac{b^5 (N-4) \left(21 N^4-160 N^3+556 N^2-960 N+648\right)}{16 w^{10}}
   + \dots
   \label{solveq}
   \end{eqnarray}
which leads to the following double-logarithmic asymptotics of the elastic amplitude:
\begin{equation}
{\cal A}_{4,(N)}(s,t) = {\cal A}_4^{\rm Born} \left(\frac{s}{|q|^2}\right)^{-\alpha |q|^2\,\ln \frac{|q|^2}{\lambda ^2}}\,\Phi^{(N)} (\xi )\,,
\end{equation}
where $\xi =\alpha \,|t|\,\ln^2\frac{s}{|q|^2}$ and 
\begin{eqnarray}
\Phi^{(N)} (\xi ) &=&1-\frac{(N-4)}{2}\,\frac{\xi }{2}+\frac{(N-4)}{2}(N-3)\frac{\xi^2 }{4!}\nonumber\\
    &-&\frac{(N-4)}{8}(5N^2-26N+36)\frac{\xi ^3}{6!} \nonumber\\
    &+&\frac{(N-4) }{8}   \left(7 N^3-47 N^2+118 N-108\right)\frac{\xi^4}{8!}\nonumber\\
     &-&\frac{(N-4) }{16} \left(21 N^4-160 N^3+556 N^2-960 N+648\right)\frac{\xi^5}{10!}
   + \dots
\label{ampexpand}
\end{eqnarray}
Before investigating in detail how the amplitudes behave at the light of this all-orders 
result, let us compare its two-loop truncation with other calculations in the literature. 

\subsubsection{Two-loop truncation and comparison with $N=4,5,6,8$ results}

In the notation of Eq.~(\ref{eqnnotation}), it is well-known that the complete 
two-loop amplitudes can be written as the sum of two pieces:
\begin{eqnarray}
{\cal M}_4^{(2),N} &=& m_4^{(2),N} +\frac{1}{2}  \left({\cal M}_4^{(1),N}\right)^2,
\label{TwoLone}
\end{eqnarray}
where $m_4^{(2),N}$ is infrared finite and has been calculated by Boucher-Veronneau and 
Dixon in Ref.~\cite{BoucherVeronneau:2011qv} for $N=4,5,6$. The $N=8$ case was calculated 
earlier in Ref.~\cite{Dunbar:1994bn}. We will now present these infrared finite remainders in the Regge limit and with leading double-logarithmic accuracy.  First, it is convenient to write the exact amplitudes in the form
\begin{eqnarray}
{\cal A}_{4,(N)} &=&{\cal A}_4^{\rm Born}
 \left(\frac{-t}{\mu^2}\right)^{\alpha t \left(\ln{\left(\frac{s}{-t}\right)}
 +i \pi \left(\frac{s}{t}\right) \right)} \nonumber\\
&\times& \Bigg\{1+ \left(\frac{N-4}{2}\right) \left(\frac{\alpha \,  t}{2}\right) \ln^2 \left(\frac{s}{-t}\right)  \nonumber\\
&&\hspace{1cm}+ \frac{1}{2} \left(\frac{N-4}{2}\right)^2 \left(\frac{\alpha \,  t}{2}\right)^2 \ln^4 \left(\frac{s}{-t}\right) 
+  m_{4, {\rm DL}}^{(2),N} + \dots \Bigg\},
\end{eqnarray}
where we have exponentiated the infrared divergent terms and singled out the non-exponentiating double-logarithmic contributions. The latter contain those pieces related to the 
one-loop result and its square, following the last term of Eq.~(\ref{TwoLone}), and the 
$m_{4, {\rm DL}}^{(2),N}$ contribution, which can be read off Ref.~\cite{BoucherVeronneau:2011qv}:
\begin{eqnarray}
m_{4, {\rm DL}}^{(2),N=4} &=& 0  \left(\frac{\alpha \, t}{2}\right)^2 \ln^4{\left(\frac{s}{-t}\right)} ~~~\Longrightarrow ~~~0,\\
m_{4, {\rm DL}}^{(2),N=5} &=& \frac{1}{24}  \left(\frac{\alpha \, t}{2}\right)^2 \ln^4{\left(\frac{s}{-t}\right)}
~\Longrightarrow ~~\frac{1}{6} \left(\frac{\alpha \,  t}{2}\right)^2 \ln^4\left(\frac{s}{-t}\right),\\
m_{4, {\rm DL}}^{(2),N=6} &=& 0  \left(\frac{\alpha \, t}{2}\right)^2 \ln^4{\left(\frac{s}{-t}\right)} 
~~~\Longrightarrow ~~\frac{1}{2}  \left(\frac{\alpha \,  t}{2}\right)^2 \ln^4\left(\frac{s}{-t}\right),\\
m_{4, {\rm DL}}^{(2),N=8} &=& -\frac{1}{3}  \left(\frac{\alpha \, t}{2}\right)^2 \ln^4{\left(\frac{s}{-t}\right)} 
\Longrightarrow ~~\frac{5}{3}  \left(\frac{\alpha \,  t}{2}\right)^2 \ln^4\left(\frac{s}{-t}\right).
\end{eqnarray}
At the right hand side of these expressions we have written the final double-logarithmic 
contribution to the amplitude. It is important to note that these results are in complete agreement 
with the two-loop truncation of our prediction for any $N$ in Eq.~(\ref{ampexpand}). Higher 
order terms for different supergravities or Einstein-Hilbert gravity can be obtained from it. This 
should serve as a useful test of multi-loop calculations of four-graviton amplitudes. As an 
example, in ${\cal N}=8$ SUGRA we obtain
\begin{eqnarray}
{\cal A}_{4,(N=8)} &=& {\cal A}_4^{\rm Born}
 \left(\frac{-t}{\mu^2}\right)^{\alpha t \left(\ln{\left(\frac{s}{-t}\right)}
 +i \pi \left(\frac{s}{t}\right) \right)} \nonumber\\
&&\times\Bigg\{1
+ 2 \left(\frac{\alpha \,  t}{2}\right) \ln^2 \left(\frac{s}{-t}\right) 
+\frac{5}{3}  \left(\frac{\alpha \,  t}{2}\right)^2 \ln^4 \left(\frac{s}{-t}\right) 
 \nonumber\\
   &&\hspace{.5cm}+\frac{37}{45} \left(\frac{\alpha \,  t}{2}\right)^3  \ln^6 \left(\frac{s}{-t}\right) 
   +\frac{353}{1260}  \left(\frac{\alpha \,  t}{2}\right)^4 \ln^8 \left(\frac{s}{-t}\right) \nonumber\\
   &&\hspace{.5cm}+\frac{583}{8100} 
    \left(\frac{\alpha \,  t}{2}\right)^5 \ln^{10} \left(\frac{s}{-t}\right)  + \dots\Bigg\}.
\end{eqnarray}
Now we turn to study the high energy asymptotic behaviour of the resummed amplitudes.

\subsubsection{Resummed supergravity amplitudes at high energies}

In terms of double-logarithmic contributions the simplest amplitude is that of ${\cal N}=4$ 
SUGRA since their contribution adds to zero and we have the pure Regge asymptotic
behaviour
\begin{equation}
{\cal A}_{4,(N=4)}(s,t) = {\cal A}_4^{\rm Born} \left(\frac{s}{|q|^2}\right)^{-\alpha |q|^2\,\ln \frac{|q|^2}{\lambda ^2}}\,.
\end{equation}
In the case of ${\cal N}=6$ SUGRA Eq.~(\ref{evoleq}) can be solved in the form
\begin{eqnarray}
f^{(N=6)}_\omega &=& \int _0^\infty d\,z\,e^{-z}\,e^{-\frac{z^2b}{2\,\omega ^2}} 
,
\label{N6}
\end{eqnarray}
obtaining the following result for the amplitude
\begin{equation}
{\cal A}_{4,(N=6)} (s,t) = {\cal A}_4^{\rm Born}\,\left(\frac{s}{|q|^2}\right)^{-\alpha |q|^2\,\ln \frac{ |q|^2}{\lambda ^2}}
\,\exp \left(-\frac{\alpha \,|q|^2}{2}\ln ^2\frac{s}{|q|^2}\right).
\end{equation}
In the general case with arbitrary $N$ it is useful to introduce the new function $y(x)$ and the new variable $x$ according to the definitions
\begin{equation}
f_\omega^{(N)} =\frac{2x}{6-N}\,y^{(N)}(x)\,,\,\,x=\frac{\omega}{\sqrt{b}},
\end{equation}
to reduce our Eq.~(\ref{evoleq}) to the Riccati equation
\begin{equation}
{y^{(N)}}'(x)+{y^{(N)}}^2(x)-x\,{y^{(N)}}+\frac{6-N}{2}=0\,.
\end{equation}
By introducing the new function $\Psi_{(N)} (x)$ as follows
\begin{equation}
y^{(N)} =\frac{d}{d\,x }\,\ln \left(e^{\frac{x^2}{4}}\Psi_{(N)} (x)\right)
\end{equation}
we obtain for it a linear Schr\"{o}dinger equation, {\it i.e.}
\begin{equation}
\left(-\frac{d^2}{d x ^2}+\frac{N-7}{2}+\frac{x ^2}{4}\right)\Psi_{(N)} (x ) =0 \,.
\label{Schroed}
\end{equation}

For even values of $N$ we have the following simple solutions for this equation:
\begin{eqnarray}
\Psi_{(N=8)}^{(1)}(x) &=& e^{\frac{x^2}{4}}\,,\\
\Psi_{(N=6)}^{(1)}(x) &=& e^{-\frac{x^2}{4}}\,,\\
\Psi_{(N=4)}^{(1)}(x) &=& x\,e^{-\frac{x^2}{4}}\,, \\
\Psi_{(N=2)}^{(1)}(x) &=& (1-x^2)\,e^{-\frac{x^2}{4}}\,,\\
\Psi_{(N=0)}^{(1)}(x) &=& x\,(3-x^2)\,e^{-\frac{x^2}{4}}\,.
\end{eqnarray}
The above solutions for $N=4,2,0$ are physical and therefore it is possible to calculate 
the following partial wave $f_\omega^{(N)}$ for these cases
\begin{eqnarray}
\frac{f^{(N=4)}_{\omega}}{\omega} &=& \frac{1}{\omega}\,, \label{N4pp}\\
\frac{f^{(N=2)}_{\omega}}{\omega} &=&
\frac{1/2}{\omega +\sqrt{b}}+\frac{1/2}{\omega -\sqrt{b}}\,,\label{N2pp}\\
\frac{f^{(N=0)}_{\omega}}{\omega} &=& 
\frac{1/3}{\omega}+\frac{1/3}{\omega +\sqrt{3b}}+\frac{1/3}{\omega -\sqrt{3b}}\,.
\label{N0pp}
\end{eqnarray}
These poles in the $\omega$-plane lead to the following double-logarithmic asymptotic behavior of the corresponding scattering amplitudes:
\begin{eqnarray}
{\cal A}_{4,(N)} (s,t) &=& {\cal A}_4^{\rm Born}\,\left(\frac{s}{|q|^2}\right)^{-\alpha |q|^2\,\ln \frac{ |q|^2}{\lambda ^2}}\,r^{(N)}(s,t),\\
r^{(N)} (s,t) &=& \int_{\delta-i \infty}^{\delta + i \infty} \frac{d \omega}{ 2 \pi i} 
\left(\frac{s}{-t}\right)^\omega \frac{f^{(N)}_\omega}{\omega},
\label{rN}
\end{eqnarray}
where
\begin{eqnarray}
r^{(N=4)}(s,t) &=&1\,,\\
r^{(N=2)}(s,t) &=& \frac{1}{2} \left(\left(\frac{s}{|q|^2}\right)^{\sqrt{\alpha |q|^2}}+\left(\frac{s}{|q|^2}\right)^{-\sqrt{\alpha |q|^2}}\right)\,,\\
r^{(N=0)}(s,t) &=& \frac{1}{3}\left(1+\left(\frac{s}{|q|^2}\right)^{\sqrt{3\alpha |q|^2}}+\left(\frac{s}{|q|^2}\right)^{-\sqrt{3\alpha |q|^2}}\right)\,.
\end{eqnarray}

The second solution of the Schr\"{o}dinger equation for even $N$ can be constructed with
the use of the Wronskian for two independent solutions
\begin{equation}
\Psi_{(N)}^{(1)}\frac{d}{d x}\,\Psi _{(N)}^{(2)}-\Psi_{(N)} ^{(2)}\frac{d}{d x}\,\Psi_{(N)} ^{(1)} 
= {\rm constant} \,.
\end{equation}
By integrating it we obtain the physical solution for ${\cal N}=8$ supergravity:
\begin{equation}
\Psi_{(N=8)} (x) = e^{\frac{x^2}{4}}\,\int _x^\infty dz\,e^{-\frac{z^{2}}{2}}=
e^{-\frac{x^2}{4}}\,\int_0^\infty dy \,e^{-\frac{y^{2}}{2}}e^{-x\,y}\,,
\end{equation}
which is proportional to the probability integral ${\rm erfc} (\sqrt{2}x)$.
We can then obtain the corresponding scattering amplitude in the form
 \begin{equation}
r^{(8)}(s,t)= -\int _{a-i\infty}^{a +i\infty}\frac{d\,x}{2\pi i }\,
\left(\frac{s}{|q|^2}\right)^{x \sqrt{b}} \,
\frac{d}{d\,x }\,\ln \int_0^\infty dy \,e^{-\frac{y^{2}}{2}}e^{-x\,y}\,.
\end{equation}

Generally, the physical solution of the Schr\"{o}dinger Eq.~(\ref{Schroed}) can be expressed in
terms of the parabolic cylinder function
\begin{equation}
\Psi_{(N)} (x)=D_{\frac{6-N}{2}}(x)\,,\,\,D_\nu (x)=\frac{e^{-\frac{x^2}{4}}}{\Gamma (-\nu)}\,
\int_0^\infty \frac{dy}{y^{\nu +1}} \,e^{-\frac{y^{2}}{2}}e^{-x\,y}\,.
\label{parabol}
\end{equation}
Therefore
\begin{equation}
\frac{f^{(N)}_{\omega}}{\omega}=\frac{2}{6-N}\,\frac{1}{\sqrt{b}}\,\frac{d}{d\,x }\,
\ln \left(\int_0^\infty  \frac{dy}{y^{\frac{8-N}{2}}}\,e^{-\frac{y^{2}}{2}}e^{-x\,y}\right)\,,\,\,x=\frac{\omega}{\sqrt{b}}\,.
\end{equation}
The integral over $y$ is convergent for $\nu <0$. For example, for $N=7$ we have
\begin{equation}
\frac{f^{(7)}_{\omega}}{\omega}=-\frac{2}{\sqrt{b}}\,\frac{d}{d\,x }\,
\ln \left(\int_0^\infty  \frac{dy}{\sqrt{y}}\,e^{-\frac{y^{2}}{2}}e^{-x\,y}\right)\,,\,\,x=\frac{\omega}{\sqrt{b}}\,.
\end{equation}

In the general case with arbitrary $\nu$ we can choose the integration contour $L$ in the complex plane $y$ which
goes from $y=+\infty$, surrounds   the
singular point $y=0$ and returns again to $y=+\infty$:
\begin{equation}
\frac{f^{(N)}_{\omega}}{\omega}=\frac{2}{6-N}\,\frac{1}{\sqrt{b}}\,\frac{d}{d\,x }\,
\ln \left(\frac{1}{2\pi i}\int_L \frac{dy}{(-y)^{\frac{8-N}{2}}}\,e^{-\frac{y^{2}}{2}}e^{-x\,y}\right)\,,
\,\,x=\frac{\omega}{\sqrt{b}}\,.
\label{fN}
\end{equation}
In particular, for $N\rightarrow 6$ we obtain
\begin{equation}
\frac{f^{(6)}_\omega}{\omega} =\frac{1}{\sqrt{b}}\int_0^\infty  dy\,e^{-\frac{y^{2}}{2}}e^{-x\,y}=
\frac{1}{\omega}\,\int _0^\infty d\,z\,e^{-z}\,e^{-\frac{z^2b}{2\,\omega ^2}},
\end{equation}
which is in agreement with Eq.~(\ref{N6}). After differentiating over $x$ in Eq.~(\ref{fN}) and taking $N=4,2,0$ we
can also reproduce our previous results in Eqs.~(\ref{N4pp},\ref{N2pp},\ref{N0pp}). 

For odd values of $N$ and $N=8$ the function $\Psi_{(N)}(x)$ has an infinite number of zeros situated asymptotically close to the lines $\arg z=\pm \frac{3}{4}\,\pi$.
The trajectories of these Regge poles satisfy the following equation at large  $n$
\begin{equation}
{x^{(N)}}^2\approx -2(7-N)\,\ln {x^{(N)}} +2\,\pi\, e^{\pm \frac{3}{2}\,\pi i} \,(2n +1) \,,\,\, n=0,1,2,...\,.
\end{equation}
For ${\cal N}=1$ SUGRA there are three Regge poles with the real trajectories
\begin{equation}
x_1^{(N=1)} \approx -2.460\,,\,\,x_2^{(N=1)}\approx -0.452\,,\,\,x_3^{(N=1)}=1,402\,,
\end{equation}
leading to the following growth of the scattering amplitude:
\begin{equation}
\lim _{s\rightarrow \infty} \,r^{(N=1)}(s,t) \approx \frac{2}{5}\,\left(\frac{s}{|q|^2}\right)^{1,402\,\sqrt{\alpha}\,|q|}\,.
\end{equation}
In ${\cal N}=3$ SUGRA  there are two real Regge trajectories,
\begin{equation}
x_1^{(N=3)} \approx -1.747\,,\,\,x_2^{(N=3)}\approx 0.5508\,,
\end{equation}
also generating a growing contribution to the amplitude:
\begin{equation}
\lim _{s\rightarrow \infty} \,r^{(N=3)}(s,t) \approx \frac{2}{3}\,\left(\frac{s}{|q|^2}\right)^{0,5508\,\sqrt{\alpha}\,|q|}\,.
\end{equation}
For the case $N=5$ there only exists one Regge pole with the real Regge trajectory
\begin{equation}
x_1^{(N=5)} \approx -0.762\,,
\end{equation}
which leads to an amplitude falling with energy 
(together with some oscilating contributions from the poles in the complex plane)
\begin{equation}
\lim _{s\rightarrow \infty} \,r^{(N=5)}(s,t) \approx 2\,\left(\frac{s}{|q|^2}\right)^{-0.762\,\sqrt{\alpha}\,|q|}\,.
\end{equation}
In the cases $N=7,8$ all the poles of $f_\omega$ are situated in the left hand side of the complex $\omega$-plane and therefore $r^{(N)}(s,t)$ here tends to zero when  $s\rightarrow \infty$.
In particular,
\begin{equation}
\lim _{s\rightarrow \infty}\,r^{(N=8)}(s,t)=
2 \,\left(\frac{s}{|q|^2}\right)^{-1.916 \,\sqrt{\alpha} |q|}\cos \left(2.8164 \,\sqrt{\alpha} |q|
\ln \frac{s}{|q|^2}\right)\,.
\end{equation}

\begin{figure}
\begin{center}
\includegraphics[width=13.4cm]{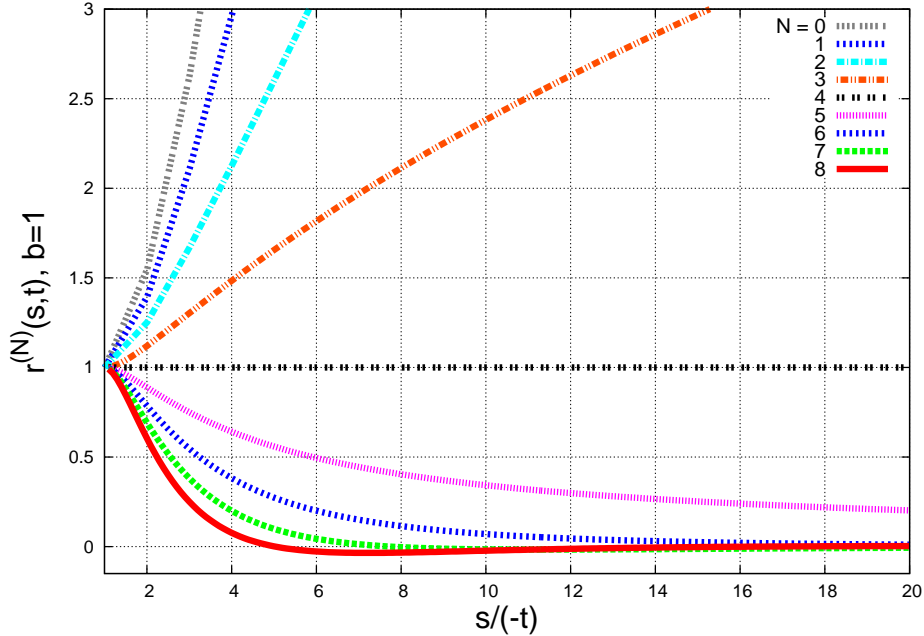}
\vspace{-1.3cm}
\end{center}
\caption{Scattering amplitude for $N=0,1,2,3,4,5, 6,7,8$.}
\label{rAllb1}
\end{figure}

\begin{figure}
\begin{center}
\includegraphics[width=12cm]{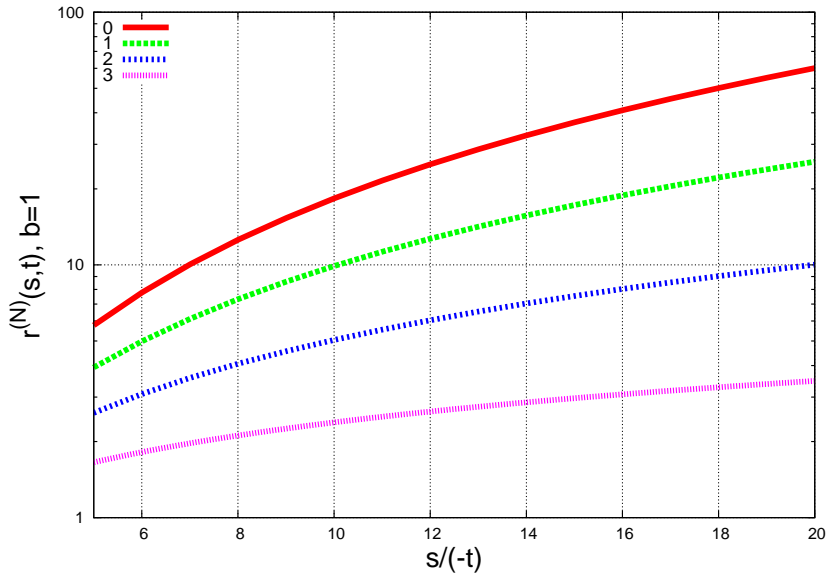}
\vspace{-1.3cm}
\end{center}
\caption{Scattering amplitude for $N=0,1,2,3$.}
\label{rN0123b1}
\end{figure}

\begin{figure}
\begin{center}
\includegraphics[width=12cm]{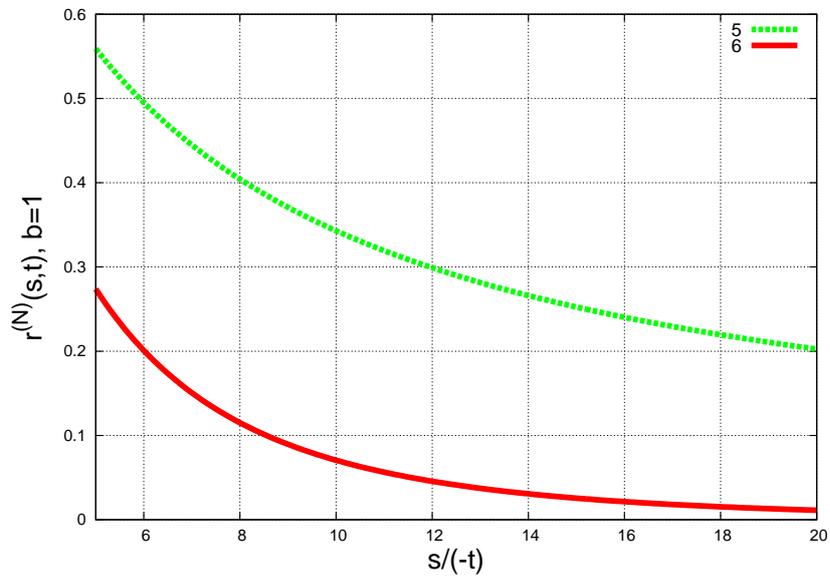}
\vspace{-1.3cm}
\end{center}
\caption{Scattering amplitude for $N=5,6$.}
\label{rN56b1}
\end{figure}

\begin{figure}
\begin{center}
\includegraphics[width=12cm]{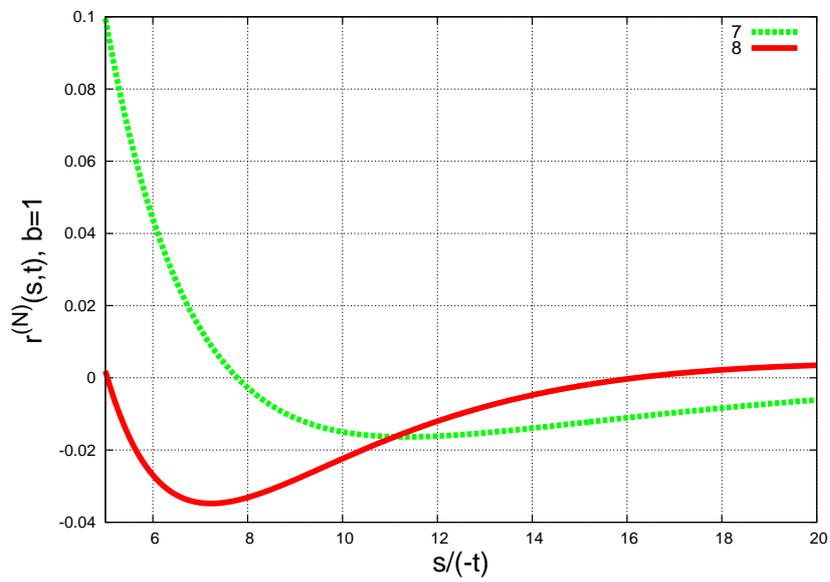}
\vspace{-1.3cm}
\end{center}
\caption{Scattering amplitude for $N=7,8$.}
\label{rN78b1}
\end{figure}

Besides these asymptotic estimates, we have performed an exact numerical analysis of the function $r^{(N)} (s,t)$ (see Eq.~(\ref{rN})). For $N=0,1,2,3,4, 5, 6, 7, 8$, in agreement with the 
previous analysis in this section, we show the energy behaviour of the scattering amplitudes 
for different supergravities  in  Fig.~(\ref{rAllb1}). 
The monotonically growing with energy solutions for $N<4$ are shown in more detail in 
Fig.~\ref{rN0123b1}. The critical solution at $N=4$ is flat with energy. The two monotonically decreasing with energy solutions for $N=5,6$ are given in Fig.~\ref{rN56b1} and the two oscillatory and decreasing with energy ones for $N=7,8$ can be seen in Fig.~\ref{rN78b1}.

It is interesting to point out  that the falling asymptotic behavior of the amplitudes at $s\rightarrow \infty$ in the ${\cal N}=5,6,7,8$ supergravities is likely to be related to good ultraviolet properties of these theories, including the possible renormalizability of  ${\cal N}=8$ SUGRA.

\section{Conclusions}

We have calculated the leading double-logarithmic in energy contributions to four-graviton scattering to all orders in the gravitational coupling. These terms are subleading with respect 
to eikonal contributions but important to understand the high energy behaviour of the scattering 
amplitudes. Our results are valid for any supergravity as well as for Einstein-Hilbert gravity. 
We have used infrared evolution equations which take into account both ladder and non-ladder topologies. The truncation of our resummation to two loops is in exact agreement with recent  calculations in the literature for ${\cal N} = 4, 5, 6, 8$ supergravities. Our results show a growth 
with energy for the amplitudes when ${\cal N}<4$, a critical invariance with the energy for ${\cal N}=4$, and an asymptotic approach to zero when ${\cal N}>4$.

\section*{Acknowledgments} 

ASV acknowledges partial support from the European Comission under contract LHCPhenoNet (PITN-GA-2010-264564), the Comunidad de Madrid through Proyecto HEPHACOS ESP-1473, and MICINN (FPA2010-17747). He also would like to thank the hospitality of the CERN PH-TH Unit, where this work started during a summer visit in 2010.

\end{document}